\documentclass[aps,prb,showpacs,twocolumn,superscriptaddress]{revtex4}
\usepackage{epsfig}

\begin{document}
\title
{Spin-polarized current oscillations in diluted magnetic semiconductor
multiple quantum wells}

\author{Manuel B\'ejar}
\affiliation{Instituto de Ciencia de Materiales de Madrid (CSIC),
Cantoblanco, 28049 Madrid, Spain}

\author{David S\'anchez}
\affiliation{Instituto de Ciencia de Materiales de Madrid (CSIC),
Cantoblanco, 28049 Madrid, Spain}
\affiliation{D\'epartement de Physique Th\'eorique,
Universit\'e de Gen\`eve, CH-1211 Gen\`eve 4, Switzerland}

\author{Gloria Platero}
\affiliation{Instituto de Ciencia de Materiales de Madrid (CSIC),
Cantoblanco, 28049 Madrid, Spain}

\author{A.H. MacDonald}
\affiliation{Department of Physics, The University of Texas at Austin,
Austin, Texas 78712}

\date{\today}

\begin{abstract}

We study the spin and charge dynamics of electrons in n-doped II--VI semiconductor
multiple quantum wells when one or more quantum wells are doped with Mn.
The interplay between strongly nonlinear inter-well charge transport and
the large tunable spin-splitting induced by exchange interactions
with spin-polarized Mn ions produces interesting new spin dependent features.
The tunneling current between quantum wells can be strongly
spin polarized and, under certain conditions, can develop
self sustained oscillations under a finite dc voltage.
The spin polarization oscillates in both magnetic and nonmagnetic
quantum wells and the time average in magnetic wells
can differ from its zero-voltage value.
Our numerical simulations demonstrate that the amplitude of the spin
polarization oscillations depends on the distribution of magnetic wells within the sample.
We discuss how the spin polarized current and
the spin polarization of the quantum wells can be tailored experimentally.

\end{abstract}

\maketitle

\section{Introduction}

Current information technology devices are based on the manipulation of electrical
charge flow using electric fields.  Recent technological developments~\cite{pri98}
that exploit magnetotransport effects in ferromagnetic metals have
increased interest in exploring other ideas. Effects based on the
electron spin degree of freedom can be used either to
improve device functionality, or to create radically new devices that either implementnew processing algorithms, as in quantum computing,
or that are based on new physical principles, as in spin-FETS.\cite{dat90}
The extreme sensitivity to external magnetic fields\cite{bai88} that
has been exploited in ferromagnetic metals is ultimately a consequence of the
collective behavior of many electronic spins that follows from
long-range ferromagnetic order.
Ferromagnetic semiconductors are important because their magnetic
properties can in principle be engineered by doping or by adjusting gate
voltages.  Considerable progress has been recently made in manipulating the
ferromagnetism that occurs in a number of common III-V compound semiconductors
when they are doped with the magnetic element Mn.\cite{fur88,ohn99,ohn00}
Although they are not normally ferromagnetic, II-VI semiconductors
doped with Mn, also respond strongly to external magnetic fields and have
properties\cite{smo97,ber00,fer01} that are similar in many respects.
The transport properties of Mn-based heterostructures have been studied
\cite{eg} including miniband transport in
strongly-coupled superlattices doped with Mn.\cite{or}

In a recent paper\cite{san02} three of us analyzed
the nonlinear transport properties of II-VI semiconductors with weakly-coupled
highly n-doped quantum-wells in which
one or more of the quantum wells is doped with Mn.
The interplay between electronic spin, charge accumulation, and
resonant interwell tunneling effects that takes place in these systems
has been shown to alter the formation of stable electric field domains
and to produce hysteretic behavior in both spin and charge
distributions. Spin polarization is induced in both magnetic and nonmagnetic
quantum wells when driven by strong dc electric fields.

In nonmagnetic highly n-doped semiconductor weakly coupled multiple-quantum-wells,
it is well known that dc electrical transport is dominated by the formation of electric field domains.
This effect is reflected in the nonlinear current-voltage characteristics\cite{wac02,agu97}
which present a sawtooth structure that arises from the interplay
between electron--electron interactions
and resonant intersubband tunneling between neighboring quantum wells.
When the carrier density is below a critical value, however, the formation of
electric field domains is not stable and the non linear transport properties change
drastically in comparision with the highly n-doped systems.
Intermediate n-doped systems have very rich behavior.
The time-averaged current--voltage
$J$--$V$ characteristics presents flat plateaus.
Real-time measurements show that within a plateau the
current has an undamped oscillatory dependence on time.\cite{kas97}
These self-sustained electronic current
oscillations\cite{san99,kas97} come from the dynamics of the domain
wall separating electric field domains and persist even
at room temperature, making these devices
promising candidates for microwave generation\cite{kas97} with frequencies
that extend from kHz to GHz.

In this paper we explore new features
that occur in 
weakly coupled quantum wells {\it that support self-sustained oscillations} 
when they are doped with magnetic impurities.   In Section II we briefly 
review the model that we use for incoherent transport between quantum wells 
which can contain magnetic impurities.   In Section III we summarize and discuss
the results we have obtained by solving this model numerically for some 
representative circumstances.  Finally, in Section IV we present our conclusions. 

\section{Model}

Our theory of transport through diluted magnetic semiconductor
 multiple-quantum-wells is built on
(i) a theory for the tunneling current between two spin-polarized two-dimensional
electron gases (2DEGs);
(ii) a continuity equation that accounts for relaxation of nonequilibrium spin populations;
(iii) a relationship between the up and down chemical potentials and their densities; 
(iv) a mean-field theory for the interaction between  2DEG electrons and 
Mn spins; and (v) an expression to take into account the Coulomb interaction between charge 
accumulations in the quantum wells.

In weakly coupled multiple-quantum-wells, it is a good approximation to treat 
tunneling of quasiparticles between neighboring 
quantum wells by leading order perturbation theory.
We ignore interwell spin-flip processes,
so that currents are carried between wells
by the two spin subsystems in parallel.
Accordingly, the current per spin $\sigma$ from the $i$th well
to the ($i+1$)st well, $J_{i,i+1}^\sigma$,
can be properly described by a {\em Transfer Hamiltonian} 
model~\cite{agu97,bar} with a Lorentzian 
lifetime broadening of each quasiparticle's spectral density.
Our Lorentzian halfwidth, $\gamma=\hbar/\tau_{\rm scatt}$,
is energy independent, purely phenomenological, and represents the 
combined effects of interface roughness, phonons, and impurity effects.
Spin relaxation procceses are not included in $\gamma$.
The scattering times in quantum wells are shorter than any other time
scale of the problem ($\tau_{\rm scatt}\simeq 0.1-1$~ps);
we therefore assume that the electrons in each well reestablish local
equilibrium between
succesive tunneling events and that their temperature is that of the lattice.
The exchange interaction that couples $s$ conduction band 
electron spins and $d$ Mn local moments
is ferromagnetic\cite{fur88}
and favors parallel alignment of the local moment $S$ and band electron spins $s$:
${\cal H}_{\rm int}^{sd}=J_{sd} \sum_{I} \vec{S}_I \cdot \vec{s}(\vec{r}_I) \,,$
where the sum is extended over the positions $I$ of the magnetic impurities and
$J_{sd}$ is the exchange integral (which we take as a constant).

When the mean-field and virtual crystal approximations are employed,\cite{fur}
the effect of the exchange interaction is to make the subband energies spin-dependent
in those quantum wells that contain Mn ions:
$E_{j}^\sigma = E_{j} - s \Delta$, where
$\Delta \equiv\ 2J_{sd}N_{\rm Mn}SB_S
( g\mu_BBS/k_B T_{\rm eff})$,\cite{fur88}
and $s=+1/2(-1/2)$ for $\sigma=\uparrow$($\downarrow$).
We must also take into account the microscopic processes that
permit the quasiparticle system to bring its spin-subsystems into
equilibrium within each quantum well.
Slow conduction band spin relaxation\cite{fla00}
makes the effects we discuss stronger
and is an important motivation for this study. Relaxation times 
in excess of $1$~ns have been established experimentally\cite{kik97}
in II-VI semiconductor quantum wells without Mn.
In II-VI diluted magnetic semiconductor quantum wells these times are reduced
to tens of picoseconds (but are still larger than $\tau_{\rm scatt}$).\cite{ba}
Fermi's Golden rule leads to the following equations for 
the spin relaxation rate equation within each well:\cite{mac99}
\begin{equation}
\frac{dn_i^\sigma}{dt}=-\frac{\mu_i^\sigma-\mu_i^{\overline{\sigma}}}
{\tau_{\rm sf}}\nu_0 \,,
\label{eq-rate1}
\end{equation}
where $\overline{\sigma}$ is the spin opposite to $\sigma$,
$\tau_{\rm sf}$ is the spin-scattering time, $\nu_0$ is the 
two-dimensional density-of-states per spin, and $n_i^\sigma$ is the
spin $\sigma$ charge density in the ith-well.
For $\Delta>\mu_i^\uparrow-E_{i\,1}^\uparrow$,
Eq.~(\ref{eq-rate1}) must be modified:\cite{san02}
\begin{eqnarray}\label{eq-rate2}
\frac{dn_i^\downarrow}{dt} & = & -\frac{n_i^\downarrow}{\tau_{\rm sf}}
= -\frac{\mu_i^\downarrow-E_{i\,1}^\downarrow}{\tau_{\rm sf}}\nu_0 \,, \\
\frac{dn_i^\uparrow}{dt} & = & - \frac{dn_i^\downarrow}{dt} \,.\label{eq-rate3}
\end{eqnarray}
For large enough $\Delta$, Eq.~(\ref{eq-rate2}) leads to an equilibrium state
with full spin polarization.
The continuity equations that describe the time evolution of the partial
densities in each quantum well include both spin-relaxation and transport
currents:
\begin{equation}
\frac{dn_i^\sigma}{dt}=\frac{J_{i-1,i}^\sigma-J_{i,i+1}^\sigma}{e}
-\frac{\mu_i^\sigma-\mu_i^{\overline{\sigma}}}{\tau_{\rm sf}}\nu_0 \quad i=1,\ldots,N
\label{eq-rate}
\end{equation}
for the case $\mu_i^\uparrow-E_{i\,1}^\uparrow>\Delta$
($N$ is the number of quantum wells).
Otherwise, Eqs.~(\ref{eq-rate2}-\ref{eq-rate3}) must replace the second term on
the right-hand side of Eq.~(\ref{eq-rate}).
\begin{figure}[t]
\centerline{
\epsfig{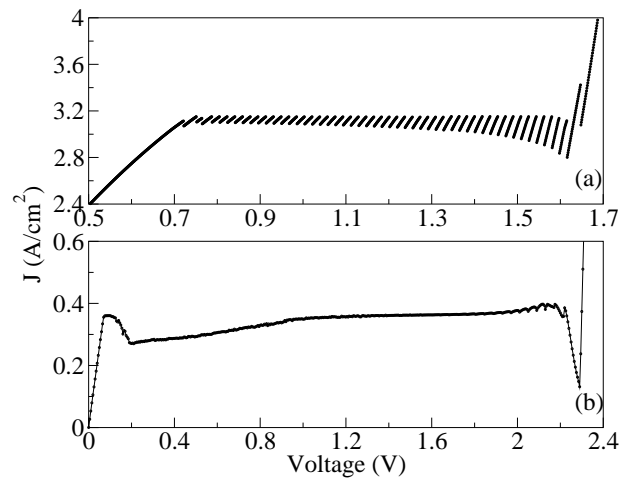}
}
\caption{$J$--$V$ curves
for a 50-well n-doped 5-nm ZnSe/10-nm Zn$_{0.8}$Cd$_{0.2}$Se system
with fractional MnSe monolayers at the 1st and 50th quantum wells.
Well doping is (a) $N_{w}=1\times 10^{12}$~cm$^{-2}$
and (b) $N_{w}=9\times 10^{10}$~cm$^{-2}$.
Contact doping is (a) $N_{c}=1.1\times 10^{12}$~cm$^{-2}$
and (b) $N_{c}=9.9\times 10^{10}$~cm$^{-2}$. Case (a) corresponds to
a highly n-doped sample. In that case, the current is stationary and
electric field domain formation takes place. It is reflected in
the sawtooth-like structure that presents J versus V. Case (b) corresponds
to an intermediate n-doped sample. In that case, the current presents
self-sustained oscillations and the time-averaged current shown in the
figure above shows a flat plateau behaviour.
}
\label{fig0}
\end{figure}

\section{Results}
In order to obtain electric field domain physics it is necessary to account for
electron-electron interactions among conduction band electrons 
at the mean-field level by solving the Poisson equation.\cite{san02}
Including a displacement current contribution,
the total current density $J(t)$ traversing the sample at time 
$t$ is $J(t)=\frac{\epsilon}{d}\frac{dV_i}{dt}+J_{i,i+1}(t)$
which is independent of $i$ and
where $J_{i,i+1}(t)=J_{i,i+1}^\uparrow(t)+J_{i,i+1}^\downarrow(t)$,
$\epsilon$ is the sample permittivity, $d$ is the multiple-quantum well
period
and $V_i$ the voltage drop between wells $i$ and $i+1$.
We model a $N=50$ n-doped ZnSe/Zn$_{0.8}$Cd$_{0.2}$Se multiple-quantum-well
system~\cite{ber00}
with well and barrier widths 10~nm and 5~nm, respectively.
Mn impurities are placed in the 1st and 50th quantum wells.
We have chosen the spin relaxation time within
the magnetic and nonmagnetic quantum wells to
be $\tau_{\rm sf}=10^{-11}$~s and 
$\tau_{\rm sf}=10^{-9}$~s, respectively.
The quantum wells and the contacts
are n-doped with $N_w$ and $N_{c}=\kappa N_w$
respectively. We take $\kappa=1.1$.
Figure~\ref{fig0} shows the $J$--$V$ characteristics for
(a) $N_{w}=1\times 10^{12}$~cm$^{-2}$ and (b) $N_{w}=9\times 10^{10}$~cm$^{-2}$.
The spin splitting is $\Delta=2$ meV in both cases. 
With these parameters, the magnetic quantum wells (if taken as isolated) are
partially polarized: 
$P=7$\% [case (a)] and $P=75$\% [case (b)], where $P_i=(n^\uparrow_i-
n^\downarrow_i)/n_i$ is the spin polarization of the $i$th-well.
As expected, Fig.~\ref{fig0}(a) exhibits a series of sharp discontinuities
in the negative differential conductance region
which are linked to the formation of static electric field domains.
Along each branch, charge
accumulates at the domain wall, forming a monopole that jumps discretely toward the emitter
as $V$ increases. When $N_w$ is lowered [see Fig.~\ref{fig0}(b)],
the branches are substituted by a plateau on which current oscillates periodically
with time. We observe undamped self-oscillations of the current in the range
$V=0.2$--$1.5$~V.
\begin{figure}[t]
\centerline{
\epsfig{file=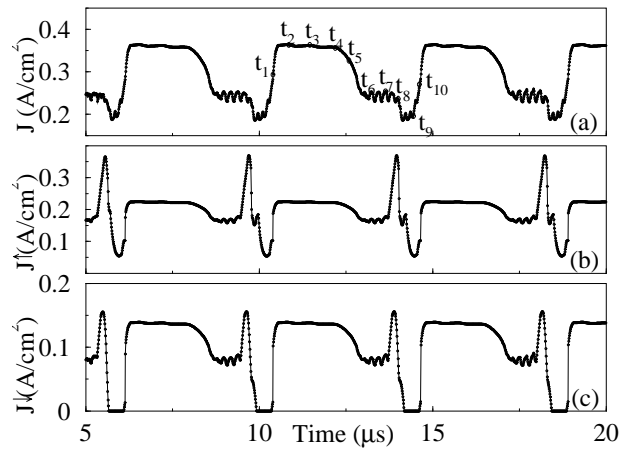,angle=270,width=0.45\textwidth,clip}
}
\caption{(a) Total time-dependent current (tunneling+displacement),
(b) spin-up, and (c) spin-down time-dependent current
at a fixed dc bias voltage $V_{\rm dc}=0.5$~V for the superlattice described 
in Fig.~\ref{fig0}(b). The current oscillations present a flat region
and overimposed spikes (see the text below).
Comparison of (b) and (c) indicates that
the current towards the collector is partially spin-up polarized.
}
\label{fig1}
\end{figure}

In Fig.~\ref{fig1} we plot
the total current (a), and the spin-up (b) and spin-down (c) currents toward the
collector as a function of time for $V=0.5$~V for the sample described in
Fig.~\ref{fig0}(b), i.e., for an intermediated n-doped sample:
$N_w=9\times 10^{10}$ and $N_c=9.9\times 10^{10}$ which presents 
self sustained oscillations in the absence of magnetic quantum wells.
The current oscillations have a period of the order of 5~$\mu$s,
{\em i.e.}, with a frequency in the kHz range.
We observe an irregular shape for the current
amplitude in all the three cases.  The current spikes that are superimposed 
on the main periodic structure reflect
discrete jumps of a domain wall separating  electric field domains 
from well to well.\cite{san99}
The current oscillations emerge from the dynamics of the domain wall
which usually consists of a charge accumulation layer (monopole) spread 
over one or two quantum wells. In nonmagnetic
multiple-quantum-wells
the dynamics of the monopole has been theoretically described
and experimentally observed.\cite{kas97} 
Lowering the contact charge doping, the domain walls lead to
traveling dipoles\cite{but79} that consist of one accumulation and
one depletion charge layer located in separated quantum wells and 
have distinct dynamics. (The dipole
charge front wave is similar to the
one responsible of the Gunn effect in bulk semiconductors).
In semiconductor multiple-quantum-wells,
monopole and dipole domain walls can 
coexist at a fixed dc bias\cite{san99} under certain conditions.
\begin{figure}[t]
\centerline{
\epsfig{file=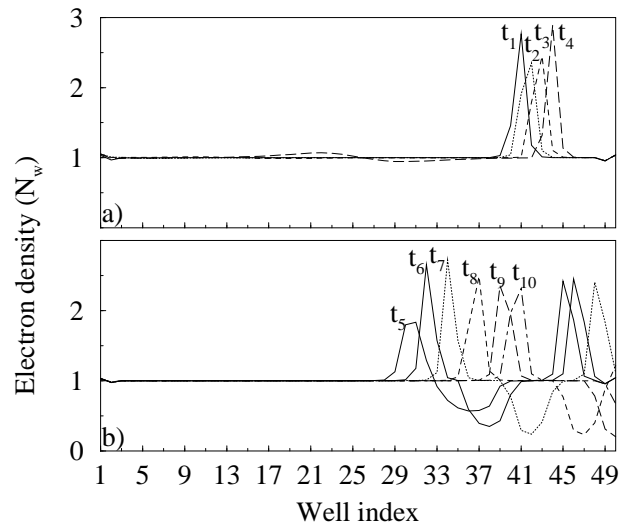,angle=270,width=0.45\textwidth,clip}
}
\caption{Local distribution of the electron density through the sample at
fixed dc bias voltage: $V_{\rm dc}=0.5$~V for the different 
times marked in Fig.~\ref{fig1}(a). At the flat region of the oscillation
in Fig.~\ref{fig1}(a) the domain wall consists of one charge accumulation
layer (monopole) which propagates just within a small part of the superlattice.
With increasing time ($t_5-t_{10}$) the domain wall configuration changes
and it becomes a dipole (one charge accumulation and one
charge depletion layer) which runs through half of the superlattice.}
\label{fig2}
\end{figure}

\begin{figure}[t]
\centerline{
\epsfig{file=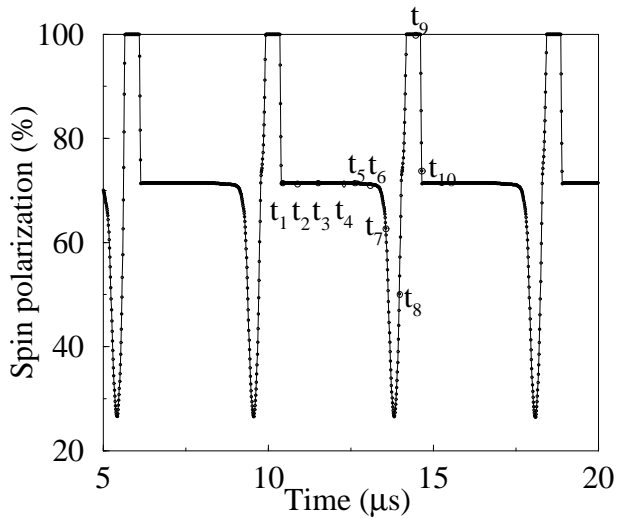,angle=0,width=0.45\textwidth,clip}
}
\caption{Spin polarization, $P$, in the quantum well 
closest to the collector
as a function of $t$ for $V_{DC}=0.5$~V. The fractional polarization of 
the isolated quantum well is $P=75$\%. Within the superlattice, in
the strongly nonlinear regime, the polarization oscillates and reaches, for a
small time window of the period, full spin-up polarization.}
\label{fig3}
\end{figure}

In Fig.~\ref{fig2}(a) we plot the charge density quantum well distribution
at the times $t_1$--$t_5$ marked in Fig~\ref{fig1}.
We observe that for the case illustrated the  domain wall
is a monopole which travels through only a part of the structure.
In Fig.~\ref{fig2}(b) the charge density at times $t_5$--$t_{10}$ is presented.
These plots illustrate an interesting feature of our results: 
the domain wall is now a dipole with both charge accumulation and depletion.
During each oscillation period the domain wall undergoes a transition
from a monopole to a dipole one and back. An exciting consequence of this 
behavior is shown if Fig.~\ref{fig3} where the polarization in the magnetic
quantum well is plotted as a function of $t$. We observe that 
the polarization reaches three different values during one oscillation:
during the intervals of dipole propagation the polarization abruptly
drops and increases up to a practically fully polarization configuration.
When the domain wall becomes a monopole the polarization remains practically
constant up to the time where the dipole is formed. The constant
value of the polarization is close to the value of the polarization
of the isolated quantum well.

\begin{figure}[t]
\centerline{
\epsfig{file=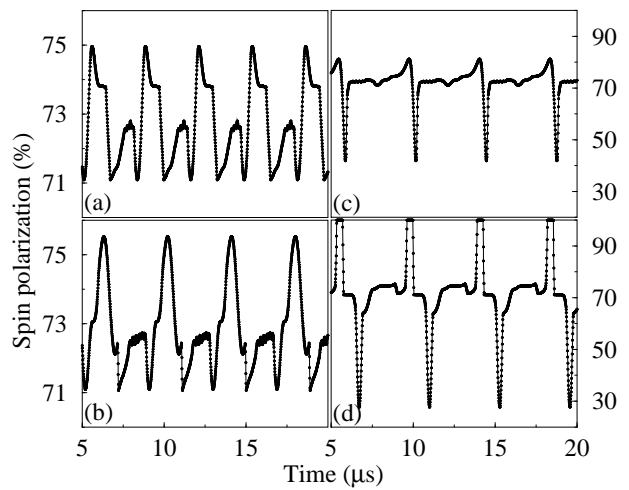,angle=270,width=0.45\textwidth,clip}
}
\caption{Spin polarization of the magnetic quantum well, 
located in the 10th (a), 20th (b), 30th (c), and 40th (d) position
at a fixed dc bias, $V_{dc}=0.5$~V. In all four cases a second magnetic quantum well
is placed at the location closest to the emitter.
The polarization corresponding to the isolated magnetic well is $P=75$\%.
The depletion charge of the dipole domain wall is centered 
around the 40th well. This depletion is responsible for the enlargement
of the spin polarization and can lead to full polarization [see (d)].}
\label{fig4}
\end{figure}

The previous observation can be explained by looking at how the  
spin polarization in the Mn-doped quantum well
influences the tunneling probability
to neighboring wells and by the strongly nonlinear transport in the 
negative differential conductance region.
The chemical potentials for spin up and down electrons
within a magnetic quantum well do not take on the same value in the stationary
limit as they would do in an isolated quantum well because of the finite
spin-flip and tunneling rates and because self-consistency
changes the charge density in the quantum wells.

The magnetic quantum well spin-polarization in the nonequilibrium
case may be related to the instantaneous local chemical potentials for majority
and minority spins:
\begin{equation}
P=\frac{\nu_{0}\Delta}{n_m}+\frac{\delta\mu_m\nu_{0}}{n_m} \,,
\label{eq-pol1}
\end{equation}
where $\delta\mu_m=\mu_m^\uparrow-\mu_m^\downarrow$.
This expression is obtained assuming the linear non-interacting 2DEG 
relationship between density and chemical potential, so that correlation
effects are not included:\cite{allan}
$n^\sigma_i=\nu_0(\mu^\sigma_i-E^\sigma_{i1})$.
The above expression for the polarization $P$
is also valid for non magnetic quantum wells. In this case, the 
spin splitting is very small and it is basically the second term of Eq.~(\ref{eq-pol1}) 
that is responsible for the induced finite polarization.

As we can see from Eq.~(\ref{eq-pol1}), increasing the density 
reduces the relative polarization. It follows that a monopole
domain wall accumulation layer in the magnetic well 
will reduce $P$ compared to the equilibrium
case. This reduction was observed
for multiple-quantum-wells with stable electric field domains
and stationary~\cite{san02} currents.
In contrast, in the self-sustained current
oscillations the magnetic well polarization can increase with respect to
the isolated case, even reaching full spin-polarization.
Comparing Figs.~\ref{fig2} and~\ref{fig3}, we note that the high polarization configuration
state occurs at $t=t_9$,
when the front wave is a dipole with one
depletion and one accumulation layer (domain walls) and the depletion
region is located at the magnetically doped quantum well.
The large polarization occurs
in part because of the local instantaneous decrease in the total
density of electrons in the magnetic well.
The frequency, amplitude and shape of the current oscillations depend
on the dynamics of the polarized charge,
which itself depends on the distribution of magnetic quantum wells
in the sample. Then tailoring the sample configuration should enable 
control of not only the frequency and amplitude of the current oscillations,
but also of the spin polarization and its time
dependence. To illustrate this we present in Fig.~\ref{fig4} results for the 
case in which two magnetic quantum wells
are placed in the system, one adjacent to the emitter and another 
at a variable position.  We observe that when the magnetic quantum well
is close to the position of the domain wall
the amplitude of the polarization oscillations increases [Fig.~\ref{fig4}(d)].
Because the polarization depends on the inverse
of the charge density, whenever the domain wall enters and 
leaves the magnetic quantum well, 
the amplitude of the magnetization oscillations increases and
$P$ oscillates between 20\% and 100\%.
As the magnetic quantum well comes closer to the collector, the polarization 
exhibits two flat regions within an oscillation.  Since the period of these
electric field domain oscillations is of the order of 5~$\mu$s,
time resolved measurements of PL in illuminated multiple quantum well samples
should allow this polarization change to be observable.

\section{Conclusion}

In closing, we have studied the spin dynamics in
multiple-quantum-well systems doped with magnetic impurities.  We predict time-dependent 
periodic oscillations of the spin polarized current and of the spin polarization
in both magnetic and nonmagnetic quantum wells. 
The interplay between nonlinearity of the current-voltage relationship 
and the exchange interaction produces new spin dependent features
which are sensitive to spin relaxation times and to the equilibrium exchange
fields in the quantum wells which can be   
tailored by adjusting external field and temperature.
These new spin dependent features
can potentially be exploited for device applications. In particular, our 
results for the oscillating spin polarized currents at the collector suggest
that these systems could be designed as spin-polarized current injection oscillators.

\section*{Acknowledgments}

This work was supported by the Spanish DGES Grant No. MAT2002-02465,
by the European Union TMR Contract FMRX-CT98-0180 and
by the European Community's Human Potential
Programme under contract HPRN-CT-2000-00144, Nanoscale Dynamics.
M. B. acknowledges the MECD for a national grant.
Research at the University of Texas was supported by the Welch Foundation
and by the Department of Energy under grant DE-FG03-02ER45958.

\end{document}